\documentclass[runningheads]{llncs}
\usepackage{hyperref}
\makeatletter
\newcommand{\printfnsymbol}[1]{%
  \textsuperscript{\@fnsymbol{#1}}%
}
\makeatother
\usepackage{marvosym}
\usepackage{mathtools}
\usepackage{graphicx}
\begin{document}
\title{CPNet: Cycle Prototype Network for Weakly-supervised 3D Renal Compartments Segmentation on CT Images}
\titlerunning{Cycle Prototype Network}
\author{Song Wang\inst{1}\thanks{equal contribution} \and Yuting He\inst{1}\printfnsymbol{1} \and Youyong Kong\inst{1,4} \and Xiaomei Zhu\inst{3} \and Shaobo Zhang\inst{5} \and Pengfei Shao\inst{5} \and Jean-Louis Dillenseger\inst{2,4} \and Jean-Louis Coatrieux\inst{2} \and Shuo Li\inst{6} \and Guanyu Yang\inst{1,4}(\Letter)}
\authorrunning{S. Wang et al.}
\institute{LIST, Key Laboratory of Computer Network and Information Integration (Southeast University), Ministry of Education, Nanjing, China
\email{yang.list@seu.edu.cn}\and
Univ Rennes, Inserm, LTSI - UMR1099, Rennes, F-35000, France \and
Dept. of Radiology, the First Affiliated Hospital of Nanjing Medical University \and
Centre de Recherche en Information Biomédicale Sino-Français (CRIBs) \and
Dept. of Urology, the First Affiliated Hospital of Nanjing Medical University \and
Dept. of Medical Biophysics, University of Western Ontario, London, ON, Canada}

\maketitle              % typeset the header of the contribution
\begin{abstract}
Renal compartment segmentation on CT images targets on extracting the 3D structure of renal compartments from abdominal CTA images and is of great significance to the diagnosis and treatment for kidney diseases. However, due to the unclear compartment boundary, thin compartment structure and large anatomy variation of 3D kidney CT images, deep-learning based renal compartment segmentation is a challenging task. We propose a novel weakly supervised learning framework, Cycle Prototype Network, for 3D renal compartment segmentation. It has three innovations: 1) A Cycle Prototype Learning (CPL) is proposed to learn consistency for generalization. It learns from pseudo labels through the forward process and learns consistency regularization through the reverse process. The two processes make the model robust to noise and label-efficient. 2) We propose a Bayes Weakly Supervised Module (BWSM) based on cross-period prior knowledge. It learns prior knowledge from cross-period unlabeled data and perform error correction automatically, thus generates accurate pseudo labels. 3) We present a Fine Decoding Feature Extractor (FDFE) for fine-grained feature extraction. It combines global morphology information and local detail information to obtain feature maps with sharp detail, so the model will achieve fine segmentation on thin structures. Our model achieves Dice of $79.1\%$ and $78.7\%$ with only four labeled images, achieving a significant improvement by about $20\%$ than typical prototype model PANet\cite{Wang_2019_ICCV}.
%\keywords{First keyword  \and Second keyword \and Another keyword.}
\end{abstract} 

\section{Introduction}
3D renal compartment segmentation is the process of extracting the 3D structure of renal cortex and medulla from abdominal CTA images, which has great significance on laparoscopic partial nephrectomy\cite{PMID:22341876,Shao2011Laparoscopic,shao2012precise,zhang2019application}. During operation, correct segmentation of renal compartments helps doctors control the proportion of nephrectomy\cite{10.1007/978-3-030-00937-3_53}, reduce the loss of renal function. Post-operatively, it assists in monitoring the recovery of renal function\cite{FICARRA2009786,PMID:19616235}, ultimately achieve the goal of reducing the cost of surgery, increasing the success rate of surgery, and providing patients with higher quality medical services.

\begin{figure}[htb]
\centering
\includegraphics[width=8cm]{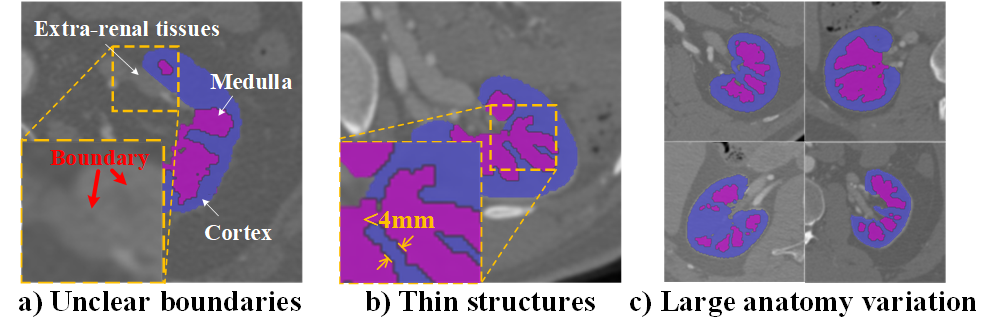}
\caption{Challenges of our renal compartment segmentation. a) Unclear boundaries between renal compartments and extra-renal tissues which make model lose the ability to extract distinguishable features. b) Thin structures of renal compartments which make feature extractor lose fine-grained features. c) Anatomy variation between images which make model sensitive to singular structures.}
\label{fig1}
\end{figure}

Deep learning has achieved remarkable success in medical image segmentation\cite{litjens2017survey,8270673}, renal tumors segmentation\cite{HELLER2021101821,PMID:33439313} and renal artery\cite{he2020dense} segmentation, but deep-learning based renal compartment segmentation on CT images is a challenging task owing to its particularity: \textbf{1) The boundary of renal compartments is not clear.} As shown in Fig.~\ref{fig1}(a), the CT values are similar between cortex, medulla and extra-renal tissues. Model will lose the ability to extract distinguishable features for compartment anatomy, therefore it is prone to over segment or under segment. \textbf{2) The structure of renal compartments is thin.} As is shown in Fig.~\ref{fig1}(b), the cortex extends into the kidney structure, entangles with the medulla to form several thin structures with unstable morphology. This makes feature extractors with large receptive fields easy to lose fine-grained features. The model trained with these features is not sensitive to small structures, thus be unable to segment the small part of renal compartments. \textbf{3) The large anatomy variation and small dataset scale.} As is shown in Fig.~\ref{fig1}(c), the renal medulla is divided into a number of random shapes. This anatomy varies between different kidneys, so fine annotation requires a lot of time for professional doctors, which limits the scale of the labeled dataset. Therefore, the labeled dataset cannot cover all compartment morphologies. Model is unable to learn generalized knowledge, will be sensitive to singular structures, and have poor segmentation capabilities for unseen morphological structures.

There is no automatic, label-efficient and high-accuracy renal compartment segmentation works on CT images being reported. Some semi-automatic works design image operation combined with manual annotation to achieve renal compartment segmentation\cite{PMID:28888170}, requiring a lot of labor costs. Deep-learning based renal compartment segmentation methods perform segmentation automatically, but the small scale of labeled dataset seriously limits their performance\cite{10.3389/fonc.2018.00215,10.1007/978-3-030-00937-3_52}.

Therefore, we proposed an automatic and label-efficient renal compartment segmentation framework Cycle Prototype Network (CPNet), which efficiently extracts renal compartments with only a few labels. It has three innovations:

1) We proposed a Cycle Prototype Learning framework (CPL) to learn consistency for generalization. It uses labels as guidance to extract features accurately and forms regularization through a reverse process to improve the generalization of the model. Feature maps are extracted under the guidance of the support label, the obtained feature vectors of the same compartment have smaller differences, and those of different compartments are more distinguishable. Prototype vectors that represent the features of compartments will be obtained by combining feature vectors of the same compartment. Prototypes are then used as templates to segment query images and train the network, forcing the feature vector extracted by the network to aggregate to the prototype vector. The feature vector of the unclear boundary deviates further from the cluster center, thus a higher penalty will be imposed. Therefore, the network will extract more discriminative boundary features. After that, the framework uses query prediction to reversely segment support images in the reverse process. This process uses the same feature extractor and prototype space to encourage the network to extract consistent class features on different images, forming a regularization thus improves the generalization ability of the model.

2) We proposed a Bayes Weakly Supervised Module (BWSM) based on cross-period prior knowledge to embed prior for pseudo label generation. Different renal compartments have different reactions on contrast agents, resulting in different performances on images of different periods. We take use of this prior to use CTA and CTU images, combined with network prediction, to obtain pseudo labels through Bayes optimization. The module first obtains noisy pseudo-labels from CTA and CTU images, which contain accurate location information, but noisy morphological information. Then it includes the network prediction as likelihood, which has relatively smooth morphological information but inaccurate location information. It uses prior knowledge Bayes theory to synthesize the two, and the obtained posterior probability weakens the error components in the two and forms a more accurate pseudo-label. Embed the posterior pseudo-label into the model indirectly expands the size of the training set. A larger training set can cover more possible compartment anatomy variations, forcing the model to reduce its attention to unstable spatial distribution features, thus improving the generalization ability of the model.

\begin{figure}[htb]
\centering
\includegraphics[width=10cm]{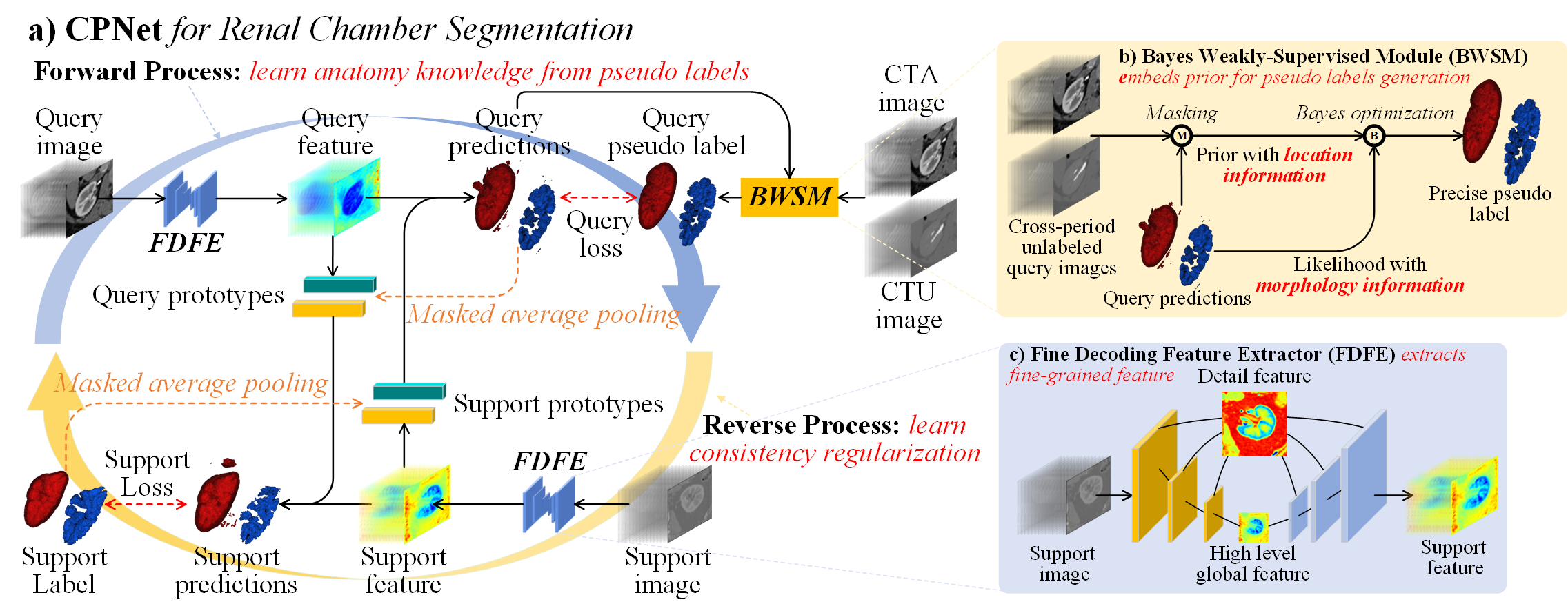}
\caption{The structure of our CPNet framework a) CPL learns consistency for generalization. It forms regularization through forward and reverse processes thus enhance robustness. b) Bayes Weakly Supervised Module. It embeds prior knowledge for pseudo label generation to enhance generalization. c) Fine Decoding Feature Extractor. It focuses on fine-grained detail information thus extract feature maps with sharp detail.}
\label{fig2}
\end{figure}

3) We proposed a Fine Decoding Feature Extractor (FDFE) that combines location information and detail information to extract fine-grained features. In the encoder-decoder stream, the decoder restores high-resolution image with the coordinate information recorded in the encoder, thus restores the global morphological information of the high-level feature maps. The cross-layer connection directly transmits the local detail information to the decoder, adds the detail feature lost in the encoder-decoder stream. Such structure combines global and local features, has better performance for segmentation tasks of renal compartments that focus on small volumes. 

\section{Methodology}

As shown in Fig.~\ref{fig2}, our CPNet uses a cycle prototype learning paradigm to efficiently implement weakly supervised segmentation of renal compartments. It has three related modules: a) The main CPL framework learns consistency with two processes. The forward process learns knowledge from pseudo labels to improve the robustness of the model. The reverse process achieves regularization and improves the generalization of the model. b) Our BWSM extracts prior knowledge and embeds pseudo-label into learning, thus improves the robustness of the model on images with large anatomy variation. c) Our FDFE combines the
global morphological features recovered by the decoder and the local detail features passed by cross-connection transmission, thus make the resulting feature maps have sharper detail information.

\subsection{CPL for consistency generalization}
\label{sec1}

\textbf{Advantages of CPL:}
1) Stronger generalization ability. The reverse process in the framework forms regularization, forcing the extracted features to meet the consistency principle, making the network robust to the noise in labels. 2) Pay more attention to the boundary area. Our framework extracts class prototypes under the guidance of label, and then imposes high penalties on boundary feature vectors that deviate from the prototype vector so that more distinguishable boundary features can be extracted.

\textbf{CPL structure for consistency regularization:}
As shown in Fig.~\ref{fig2}, our framework trains an efficient feature extraction network on weakly supervised dataset, and consists of two processes: forward process and reverse process. It first uses FDFE to extract support feature $x_s$ and query feature $x_q$ from the support image $i_s$ and query image $i_q$. In the forward process, it uses the support label $y_s$ to perform masked average pooling $M(\cdot)$ to obtain the support prototype $t_s= M(y_s\cdot x_s)$. Feature vectors in $x_q$ is classified by calculating cosine similarity $CS(\cdot)$ with $t_s$ to obtain query prediction $y'_q= CS(t_s\cdot x_q)$. Similarly, in the reverse process, it uses $y'_q$ as query label, extracts the query prototype $t_q = M(y'_q\cdot x_q)$ and predicts the support image to obtain the support prediction $y'_s = CS(t_q\cdot x_s)$.

\textbf{Forward and reverse learning process:}
We train our model through forward and reverse process. In the forward process, our model learns from query pseudo labels, so we set query loss $L_{query}$ to optimize the performance of our model. It is calculated between query prediction $y'_q$ and query pseudo-label $\hat{y}_q$, and is used to measure the robustness of the model on various query images. In reverse process, our model learns consistency regularization. If the query prediction we get in forward process is accurate, reverse process will recover the correct support label with it. Therefore, the support loss $L_{support}$ calculated between support prediction $y'_s$ and support label $y_s$ is set to measure the generalization of the model on recovering support label. Both losses are cross-entropy loss\cite{he2020dense}, the total loss of our learning process is ${L}_{total}= \theta{L}_{query} + {L}_{support}$, where $\theta$ is the query loss weight hyperparameter used to balance these losses.

\subsection{BWSM for prior embedding}

\textbf{Advantages of BWSM:}
1) Enlarges training dataset. It extracts prior knowledge from unlabeled data and embeds it into learning, which indirectly expands the scale of the training set and improves generalization. 2) Extracts accurate pseudo-labels. Prior pseudo-labels are optimized by network prediction, thus reduces the influence of noise and obtains more accurate pseudo-labels.

\begin{figure}[htb]
\centering
\includegraphics[width=7cm]{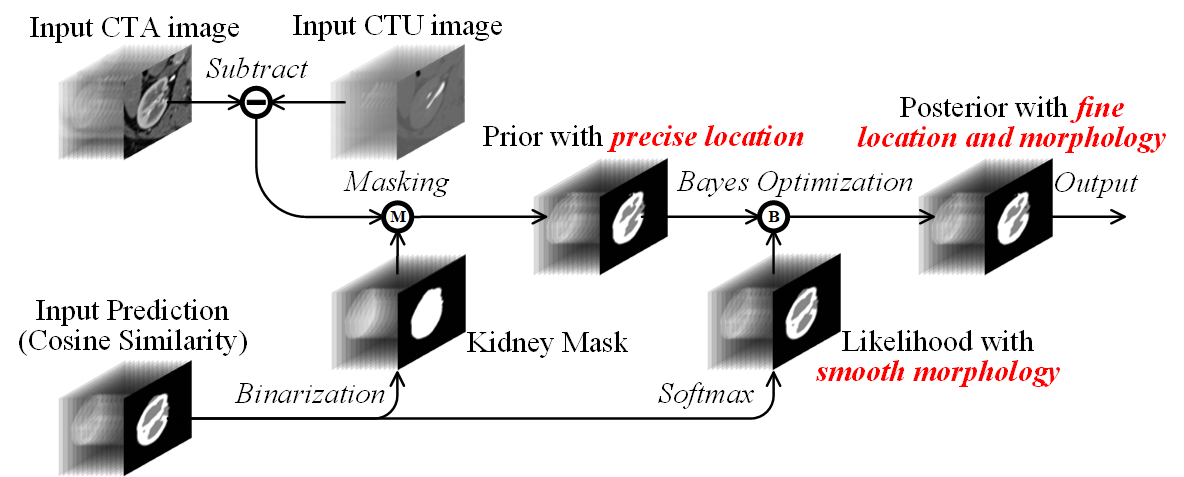}
\caption{BWSM for prior embedding. Our BWSM combines prior knowledge and network prediction to generate precise pseudo label.}
\label{fig3}
\end{figure}

\textbf{BWSM process of pseudo label generation:}
As shown in Fig.~\ref{fig3}, our BWSM has a prior knowledge extraction process and a Bayes correction process. The prior knowledge extraction process uses the different appearance of compartments in CTA and CTU images to produce a prior prediction of renal compartments. It first filters the CTA and CTU images and then subtracts them to obtain the prior feature map $f_q$, which multiplies the kidney mask $f_q$ generated from network prediction $y'_q$ to obtain the prior probability for correct and wrong predictions $p_{correct}$ and $p_{wrong}$. The Bayes correction process combines network prediction to correct the prior pseudo-label to obtain a more accurate posterior pseudo-label. The softmax probability is used as the likelihood probability $l_{correct}$ and $l_{wrong}$, then it is used to modify $p_{correct}$ and $p_{wrong}$ to obtain the posterior pseudo-label ${f}_{correct}$ and ${f}_{wrong}$. The process of Bayes correction process is as follows:
\begin{equation}
\left\{
\begin{array}{lr}
    {p}_{correct} = 1/3 + \omega\\
    {p}_{wrong} = (1 - {p}_{correct})/2\\
    {p}_{correct} = \frac{{p}_{correct}*{l}_{correct}}{{p}_{correct}*{l}_{correct}+{p}_{wrong}*{l}_{wrong}}\\
    {f}_{wrong} = \frac{{p}_{wrong}*{l}_{wrong}}{{p}_{correct}*{l}_{correct}+{p}_{wrong}*{l}_{wrong}}
\end{array}
\right.
\end{equation}
where $\omega$ is the prior probability difference hyperparameter to balance the influence of prior pseudo label.

\subsection{FDFE for fine-grained feature extracting}
\textbf{Advantages of FDFE:}
1) Emphasizes global morphology restoration. The decoder inherits the position information saved by the encoder thus performs spatial restoration more accurately. 2) Emphasizes the extraction of local detail features. The cross-layer connection transmits high-resolution features without downsampling, so the output of the network has sharper details. 3) Enhances segmentation on thin structures. A combination of morphology and detail information makes the output feature maps have sharp detail features. Such feature maps will make the model able to segment thin renal compartment structures.

\textbf{Structure of FDFE:}
As shown in Fig.~\ref{fig2} (c), our FDFE combines global information and detail information to extract fine-grained features. Global information is restored by up-pooling in the encoder-decoder stream. Detail information is retained by skip connection between convolution blocks of the same dimension. Specifically, the encoder consists of several repeated blocks. Each block contains two $3\times3\times3$ convolutional layers and a $2\times2\times2$ pooling layer, each convolutional layer is followed by a group norm layer. The decoder consists of several convolutional blocks corresponding to the decoder.

\section{Experiments and Results}

\textbf{Experiment settings: }
Our dataset is obtained by preprocessing the abdominal enhanced CT images of patients undergoing LPN surgery. The pixel size of these CT images is between $0.59mm^{2}$ to $0.74mm^{2}$. The slice thickness is fixed at $0.75mm$, and the z-direction spacing is fixed at $0.5mm$. 60 kidney ROIs of size $160\times160\times200$ were used in the research, half of which were used as the training set and the other half as the test set. 4 images in the training set and all 30 images in the test set are fine labeled. We trained our model for 20,000 iterations. In each iteration, we randomly select two supporting images for two prototypes representing two renal compartments and one query image from the corresponding data set to form an episode for training. The support and query sets in the first 2000 iterations are all from 4 labeled training images. The source of the support set in the last 18000 iterations remains unchanged. The query set is taken from the remaining 26 unlabeled training images. During the test process, the support set is still extracted from 4 labeled training images, while 30 test images work as the query set aimed to be segmented.

We use the SGD optimizer with learning rate $lr=0.001$, momentum of 0.9, and batchsize of 1. When using the Bayes algorithm to optimize pseudo-labels, we assign the prior probability difference hyperpatameter $\omega=0.05$. When using the fully-supervised data set to initialize the network, the query loss weight hyperparameter is set at $\theta=1$, and when the pseudo-label is introduced for training, the $\theta$ is changed to 0.1. We use mean Dice for medulla and cortex (Dice-M and Dice-C), and Average Hausdorff Distance (AHD-M, AHD-C)\cite{he2020dense} to parametrically measure model performance.

\begin{figure}[htb]
\centering
\includegraphics[width=9cm]{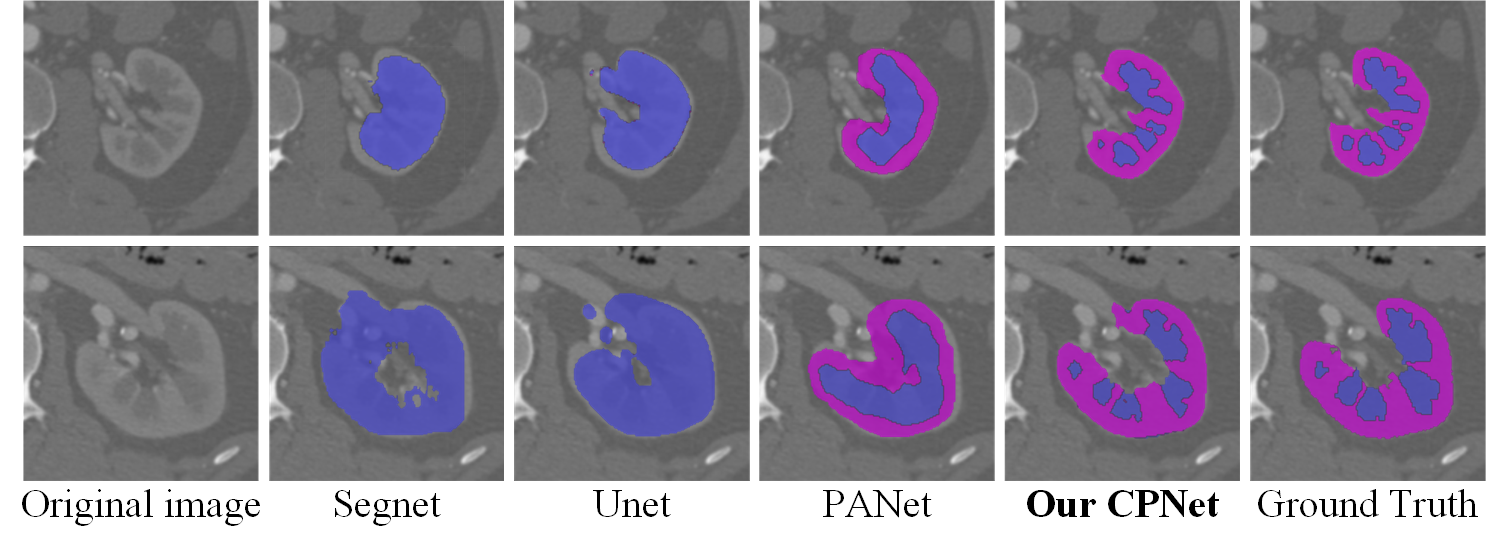}
\caption{The visual superiority of our CPNet. Segnet and Unet only obtain the rough boundary of the whole kidney. Prototypical method PANet can roughly segment renal compartments but have serious detail loss. Our CPNet learns detail information on the four labeled images better and achieves fine segmentation.}
\label{fig4}
\end{figure}

\textbf{Comparative analysis:} Our framework has the best performance compared to other methods. As shown in Tab.~\ref{tab1}, given four labeled images, our mean Dice of cortex segmentation is 78.4, and 79.1 for medulla segmentation. Some methods achieve very low performance. Segnet and Unet did not learn enough knowledge from the given four labeled images, so they cannot correctly segment renal compartments. Prototypical method PANet achieves Dice of 55.9 and 56.7. As shown in Fig.~\ref{fig4}. Segnet and Unet judge the entire kidney structure as medulla and cannot correctly segment the cortex. PANet roughly segments renal compartments, but there is serious detail loss. Our CPNet retains detail better and achieves fine segmentation.

\begin{table}[!htb]
\centering
\caption{Our method achieves best result on AHD and Dice. P are prototype methods.}
\resizebox{\textwidth}{!}{
\begin{tabular}{lllllll}
\hline
Network &Dice-M(\%)  &Dice-C(\%)  &Avg Dice(\%) &AHD-M    &AHD-C    &Avg AHD\\
\hline
% V-Net\cite{7785132} &Unable &Unable  &Unable   &Unable\\
SegNet\cite{7803544}    &65.2$\pm$5.6 &Unable  &-   &6.1$\pm$1.3    &Unable &-\\
U-Net\cite{10.1007/978-3-319-24574-4_28}&72.7$\pm$8.6  &2.2$\pm$14.0    &37.4$\pm$35.4  &4.1$\pm$1.5    &9.4$\pm$2.4    &6.8$\pm$3.4\\
(P)PANet\cite{Wang_2019_ICCV}   &55.9$\pm$9.2 &56.7$\pm$8.8 &56.3$\pm$9.0   &3.7$\pm$1.0    &3.5$\pm$1.1    &3.6$\pm$1.1\\
\hline
(P)Our CPNet-BWSM  &58.1$\pm$7.6 &59.6$\pm$6.8  &58.9$\pm$7.2 &3.3$\pm$0.9    &3.1$\pm$0.8    &3.2$\pm$0.9\\
(P)Our CPNet-FDFE  &76.5$\pm$9.0    &77.0$\pm$9.2 &76.8$\pm$9.1  &2.6$\pm$1.1    &2.6$\pm$1.1   &2.6$\pm$1.1\\
(P)Our CPNet-Total  &\textbf{78.4$\pm$9.2}    &\textbf{79.1$\pm$7.9}    &\textbf{78.7$\pm$8.6} &\textbf{1.8$\pm$0.8}    &\textbf{1.7$\pm$0.7}   &\textbf{1.7$\pm$0.8}\\
\hline
\end{tabular}}
\label{tab1}
\end{table}

\textbf{Analysis of innovations:} The results show that the use of the FDFE has a significant improvement in the effect compared with the end-to-end semantic feature extractor network (VGG16), achieves an improvement on mean Dice by about $20\%$. Our BWSM is also important for network performance. Compared with training without this module, mean Dice improves by about $2\%$.

\begin{figure}[htb]
\centering
\includegraphics[width=10cm]{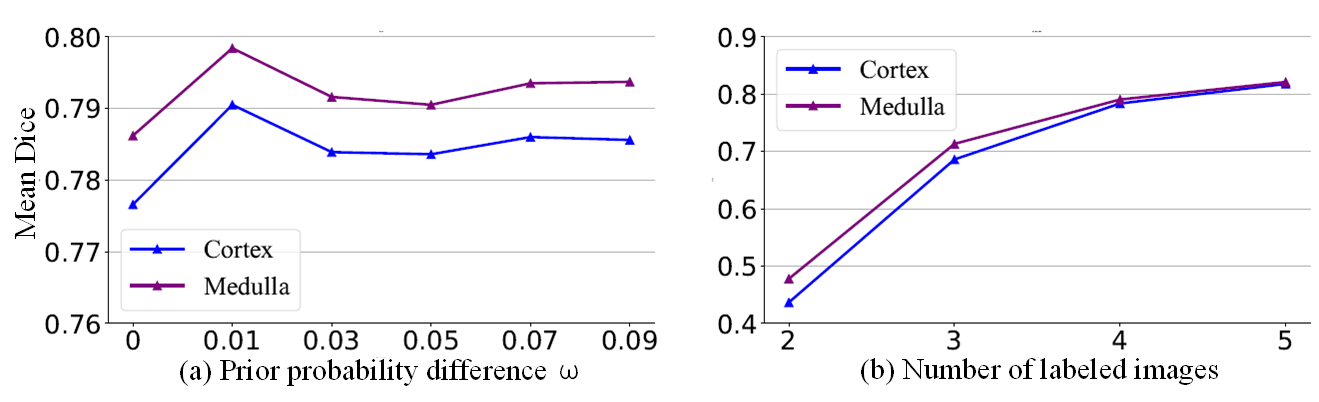}
\caption{Our hyperparameter analysis. a) Model performance increases then decreases as $\omega$ increases, achieving the best performance at $\omega=0.01$. b) Model performance increases as the number of labeled images increases. The increase speed gradually decreases and finally stabilizes.}
\label{fig5}
\end{figure}

\textbf{Hyperparameter analysis:}
\textbf{Prior probability difference $\omega$:} As shown in Fig.~\ref{fig5} (a), Model performance increases then decreases as $\omega$ increases and achieves the best outcome at $\omega=0.01$. $\omega=0$ means BWSM directly takes network prediction as the pseudo label, and high $\omega$ means BWSM directly takes prior probability as the pseudo label. Both of them are biased and need correction, so we set $\omega$ at a balanced point. \textbf{Data amount analysis:} As shown in Fig.~\ref{fig5} (b), model performance increases with the increase of the number of labels. More labels contain more information, so it is obviously better for training. In order to prove our superiority under weak supervision, we set the number of labels at a relatively small amount.

%  a) Model performance decreases as $\theta$ increases, achieving best performance at $\theta=0.1$.
% \textbf{Query loss weight $\theta$:}As shown in Fig.~\ref{fig5} (a), model performance decreases with the increase of $\theta$, and achieves best outcome at $\theta=0.1$. Higher $\theta$ means model focus more on query loss calculated between query prediction and query pseudo label. This pseudo label contains noisy prior information, so we set this weight at a limited value.

\section{Conclusion}
In this article, we propose a new automatic renal compartment segmentation framework Cycle Prototype Network on 3D CT images. The main Cycle Prototype Learning framework uses labels as guidance to extract features accurately and forms regularization through a reverse process to improve generalization. In addition, by embedding the Bayes Weakly Supervised Module into the framework, it can learn from unlabeled data autonomously, which improves the generalization of the framework. A unique Fine Decoding Feature Extractor is adopted to further strengthen the framework's capability to extract fine-grained detail features. The experiment results show that this method has obtained satisfactory segmentation accuracy and has potential for clinical application.

\paragraph{\textbf{Acknowledgments}}This research was supported by the National Natural Science Foundation under grants (31571001, 61828101, 31800825), Southeast University-Nanjing Medical University Cooperative Research Project (2242019K3DN08) and Excellence Project Funds of Southeast University. We thank the Big Data Computing Center of Southeast University for providing the facility support on the numerical calculations in this paper. 
\bibliographystyle{splncs04}
\bibliography{mybib}
\end{document}